\documentclass[fleqn,10pt]{wlscirep}
\usepackage[utf8]{inputenc}
\usepackage[T1]{fontenc}
\usepackage{hyperref}
\usepackage{algorithm2e}
\usepackage{graphicx}
\usepackage{geometry}

\SetKwComment{Comment}{/* }{ */}
\RestyleAlgo{ruled} 

\title{Understanding microbiome dynamics via interpretable graph representation learning}

\author[1,*]{Kateryna Melnyk}
\author[2]{Kuba Weimann}
\author[2]{Tim O.F. Conrad}
\affil[1]{Department of Mathematics and Computer Science, Freie Universit\"at Berlin,
        Arnimallee 6, Berlin, 14195, Germany}
\affil[2]{Zuse Institute Berlin, Takustraße 7, Berlin, 14195, Germany}

\affil[*]{Corresponding author: melnykk96@zedat.fu-berlin.de}


\begin{abstract}
Large-scale perturbations in the microbiome constitution are strongly correlated, whether as a driver or a consequence, with the health and functioning of human physiology. However, understanding the difference in the microbiome profiles of healthy and ill individuals can be complicated due to the large number of complex interactions among microbes. We propose to model these interactions as a time-evolving graph where nodes represent microbes and edges are interactions among them. Motivated by the need to analyse such complex interactions, we develop a method that can learn a low-dimensional representation of the time-evolving graph while maintaining the dynamics occurring in the high-dimensional space. Through our experiments, we show that we can extract graph features such as clusters of nodes or edges that have the highest impact on the model to learn the low-dimensional representation. This information is crucial for identifying microbes and interactions among them that are strongly correlated with clinical diseases. We conduct our experiments on both synthetic and real-world microbiome datasets.
\end{abstract}
\begin{document}

\flushbottom
\maketitle
%
%
\thispagestyle{empty}

\section*{Introduction}

\label{sec:intro}
Complex microbiome ecosystems have a strong impact on the health and functioning of human physiology. Large-scale perturbations in the microbiome constitution are strongly correlated, whether as a driver or a consequence, with clinical diseases, such as inflammatory bowel disease \cite{InflammatoryMicrobiome, InflammarotyMicrobiome1}, obesity \cite{ObesityMicrobiome2}, and some types of cancer \cite{CancerMicrobiome1, CancerMicrobiome2, CancerMicrobiome3, CancerMicrobiome4}. 

\begin{figure}[ht]
	\centering
		\includegraphics[scale=.20]{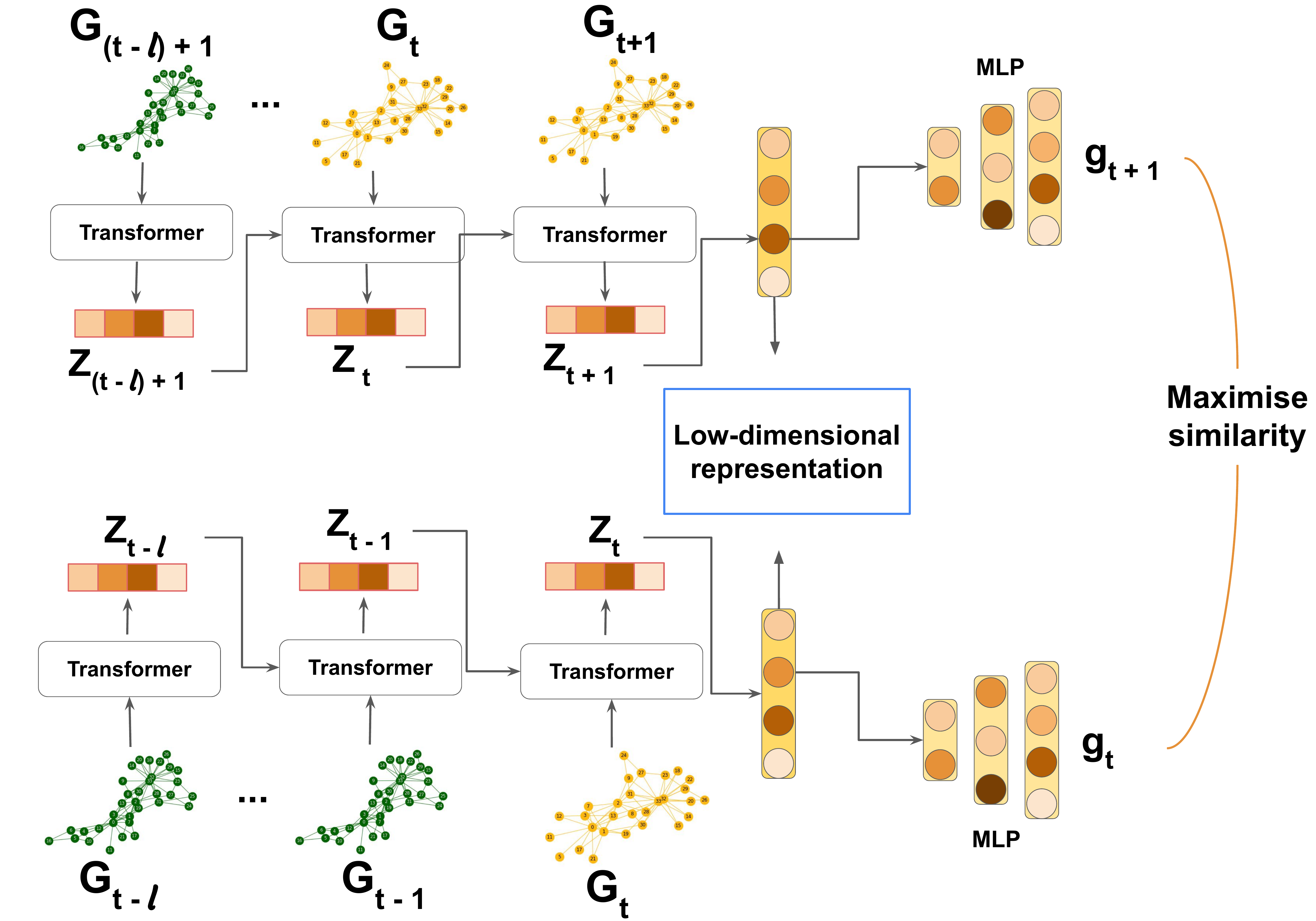}
	\caption{Overview of the method proposed in this paper.}
	\label{fig:FIG1}
\end{figure}

Many studies have been aimed at accurately differentiating the disease state and at understanding the difference in the microbiome profiles of healthy and ill individuals \cite{StatisticApproach1,MovingPicture}. However, most of them mainly focus on various statistical approaches, omitting microbe-microbe interactions between a large number of microbiome species that, in principle, drive microbiome dynamics. In addition, some studies make use of the concept of a potential landscape in physics \cite{PhysicLandscape, Shaw2019ModellingMR, Faust2015MetagenomicsMT}, giving a completely new insight into the analysis of microbiome dynamics. Namely, a healthy human microbiome can be considered as a metastable state lying in a minimum of some potential landscape. The system of time-evolving interactions of species appears to be equilibrated for a short timescale but at larger timescales a disease or other strongly impacting factors, such as antibiotic exposure, makes the system undergo transitions from one metastable state (healthy) to other metastable states (diseased).

Detecting metastable states and associated interactions of species, which undergo changes from one metastable state to others, is complicated by the high dimensionality and the compositional nature of microbiome data. Therefore, we propose a method that simplifies analysis and prediction of the large-scale dynamics of microbiome composition by projecting this system onto a low-dimensional space. First, to allow interactions between species to change over time, we represent the system as a time-evolving graph with nodes being microbes and edges being interactions between microbes. Second, we define two key components of our method: 1) the Transformer \cite{AttentionWhatYouNeed} that learns both structural patterns of the time-evolving graph and temporal changes of the microbiome system, and 2) contrastive learning that makes the model maintain metastability in a low-dimensional space. To assess the performance of our method, we apply it to the synthetic data from \cite{graphkke}, which has known underlying dynamics, and to two real-world microbiome datasets, i.e. MovingPic \cite{MovingPicture} and Cholera Infection \cite{CholeraInfOriginal}. Furthermore, we will show that it is feasible to extract topological features of the time-evolving graph which are associated with metastable states and have the highest impact on how the model learns the low-dimensional representation of the time-evolving graph with metastability. This information can help in differentiating the microbiome profile of healthy and diseased individuals.  

Our main contribution is presenting a model that learns a low-dimensional representation of the time-evolving graph with metastable behaviour in an unsupervised manner. We show in experiments that the metastability governing the time-evolving graph is preserved by the model. Through interpreting the output of the model with respect to the input, we demonstrate that it is feasible to extract topological features of the time-evolving graph, which define each metastable states.  These features can be used to identify a set of microbes that drive microbiome constitution to undergo transitions from one metastable state to others.

\section*{Related work}
We can broadly categorize methods for graph representation learning into semi-supervised or unsupervised methods and methods for static or time-evolving (dynamic) graphs. A good overview of the current state of methods for time-evolving and for static graph representation techniques can be found in \cite{Kazemi2020RepresentationLF, Barros2021ASO} and in \cite{Cui2019ASO, Zhang2020NetworkRL}, respectively. The most recent survey on both time-evolving graphs and static graphs is presented in this paper \cite{GraphRepSurvey2022}.

\paragraph{\textbf{Static graph representation.}} Representation approaches for static graphs can be classified into two categories -- those which learn the representation of nodes and those which learn the representation of sub-structures of the graphs. The first category tends to encode nodes of the graph in a low-dimensional space such that their topological properties are reflected in the new space (node2vec \cite{node2vec}, DeepWalk\cite{DeepWalk}). Most studies are focused on node representation learning, and only a few learn the representation of the whole graph (graph2vec \cite{graph2vec}).

\paragraph{\textbf{Dynamic graph representation.}} Representing time-evolving graph in the low-dimensional space is an emerging topic that is still being investigated. Among recent approaches, DynGEM \cite{Goyal2018DynGEMDE} uses the learned representation from the previous time step to initialize the current time step representation. Such initialization keeps the representation at the current time step close to the learned representation at the previous time step. The extension of the previous method is dyngraph2vec \cite{dyngraph2vec}, where authors have made it possible to choose the number of previous time steps that are used to learn the representation at the next time step. Moreover, dyngraph2vec uses recurrent layers to learn the temporal transitions in the graph. Unlike this method, we utilize the multi-head attention mechanism \cite{AttentionWhatYouNeed} to capture the temporal changes in the time-evolving graph.

Another category of methods that are successful in the graph representation learning is Graph Neural Networks. One of methods is EvolveGCN \cite{EvolveGCN} which extents graph neural network for static graph to dynamic graphs through introducing a recurrent mechanism to update the network parameters. The author focus on the graph convolutional network and incorporate a recurrent neural network to capture the dynamics of the graph. Recently, attention-based methods have been extensively proposed, and one of them is DynSAT (\cite{DySat}) that learns a dynamic node representation by considering topological structure (neighbourhood) and historical representations following the self-attention mechanism. However, one of the disadvantages of these methods for our problem is that time-evolving graphs with metastability usually consist of many time steps, and it is crucial to have a computationally efficient method in order to learn a low-dimensional representation. Another disadvantage is that all these methods capture the dynamic of nodes and, as the result, output the low-dimensional representation of nodes. 

\section*{Results}

\subsection*{Datasets}
Here, we briefly describe the datasets used to evaluate the model. Besides experiments with synthetic datasets, we show the application of our method to real-world microbiome data. Both the idea of generating synthetic dataset and the idea of pre-processing real-world datasets are explained in more details in \cite{graphkke}. An overview of the datasets used in this paper is shown in Table \ref{tab:data_stat}.

\subsubsection*{Synthetic data} 
\label{sec:synthetic_dataset}
To estimate how the proposed method can capture the dynamics of the time-evolving graph and learn a proper low-dimensional representation, we generate synthetic datasets with both understandable topological structures and temporal patterns. We use the following Stochastic Differential Equation (SDE) and a potential function to sample a trajectory with metastability based on which time-snapshot graphs $\{G_1, \dots\, G_{T}\}$ are built:


\begin{figure}[ht]
	\centering
		\includegraphics[scale=.3]{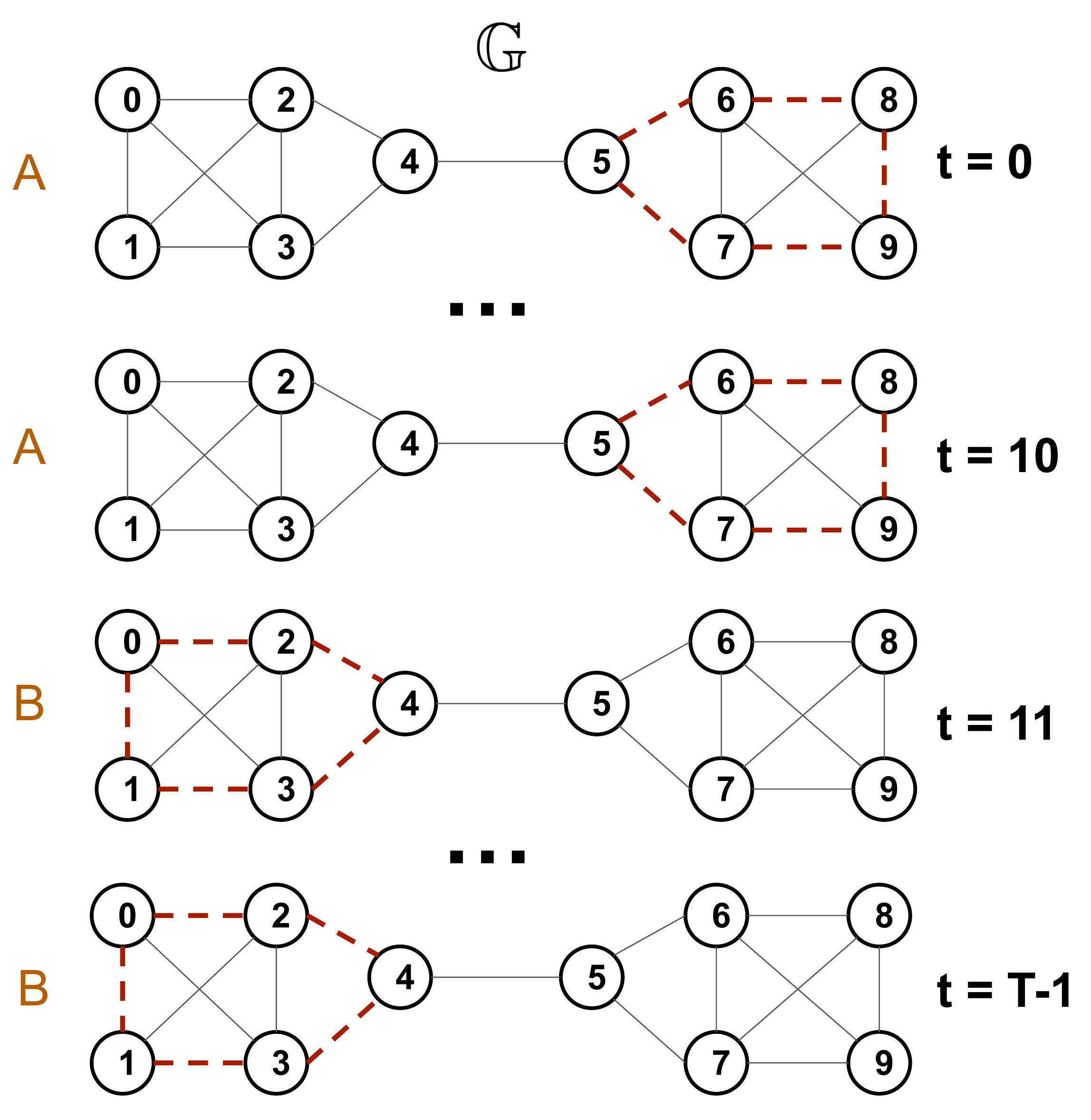}
	\caption{The example of a time-evolving graph with metastability where two states \textit{A} and \textit{B} are difficult to distinguish since they are topologically the same. Red dashed edges are removed from the time-evolving graph $\mathbb{G}$.}
	\label{fig:FIG2}
\end{figure}

\begin{equation}
    dX_t = -\nabla F(X_t)dt + \sqrt{2 \beta^{-1}}dW_t 
    \label{eq:sde}
\end{equation}
with the potential function:
\begin{equation}
    F(x) = \cos (s \arctan (x_2, x_1)) + 10\Big(\sqrt{x_1^2 + x_2^2} - 1\Big)^2,
    \label{eq:limon_potential}
\end{equation}
where \textit{s} is the number of states or wells, $X_t = (x^1_t, x^2_t)$, $W_t$ is a standard Wiener process, $\beta$ is the inverse temperature that controls the transition between states. The higher $\beta$, the less likely is the transition from one state to another. We use the potential function \eqref{eq:limon_potential} to generate pos\_3WellGraph.

For the synthetic datasets with 2 states (pos\_2WellGraph and npos\_2WellGraph), we use the following potential function, which is often called a double-well potential:

\begin{equation}
    F(x) = \frac{x^4}{4} - \frac{x^2}{2},
    \label{eq:double_potential}
\end{equation}
where $X_t = x_1$.

We further split our synthetic datasets into two categories: positional and non-positional data. Positional data means that we use the positional encoding explained in Section \ref{sec:proposed_method} before training the model. Non-positional data means that we do not use the positional encoding, as the topological structure of the time-evolving graph can be understood without providing the positional information of the node. We define these two categories to empirically show that our model does not rely on positions of nodes if the topological patterns of each state are clearly defined.

\begin{figure}[ht]
	\centering
		\includegraphics[scale=.6]{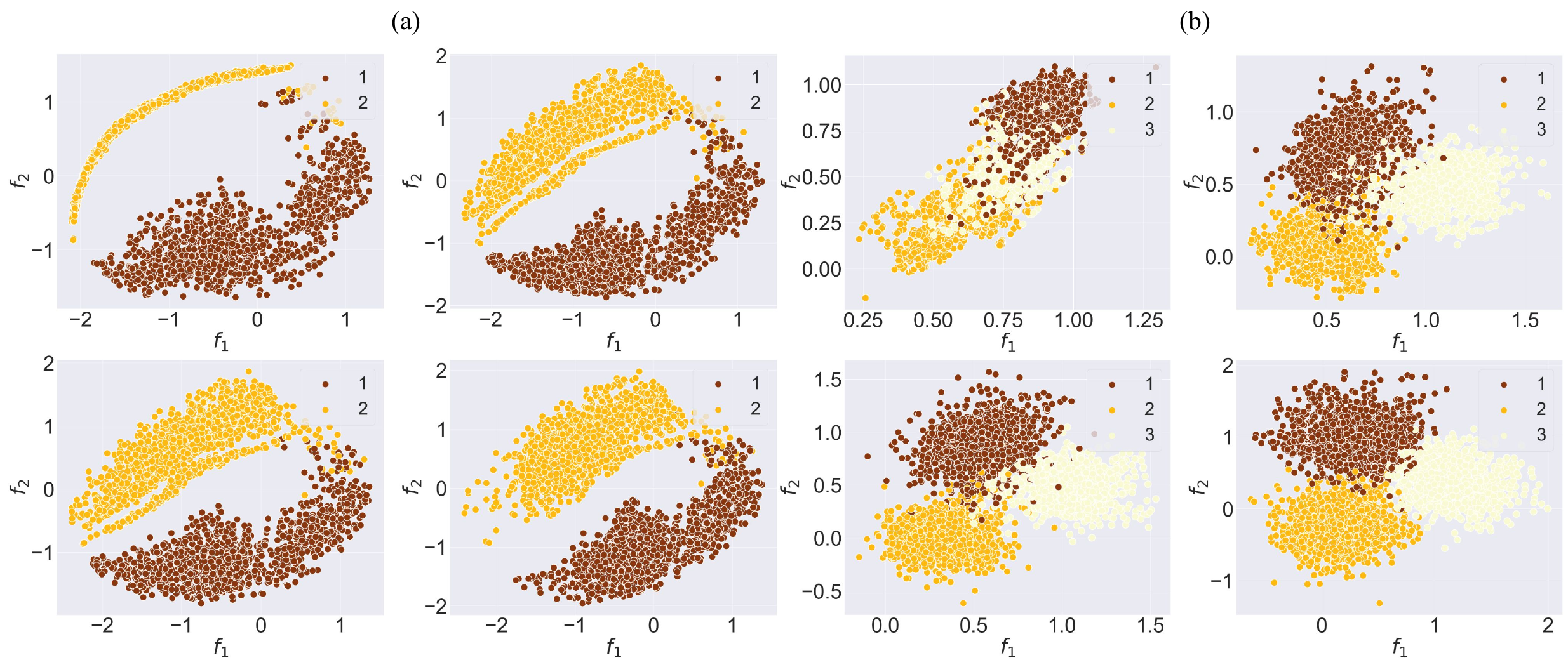}
	\caption{The evolution of the graph embedding of the time-evolving graph $\mathbb{G}$ during the training of our model on (a) npos\_2WellGraph and (b) pos\_3WellGraph. The points are colored according to ground-truth labels.}
	\label{fig:FIG3}
\end{figure}

\paragraph{\textbf{Positional data.}}  This synthetic data has a topological structure that is difficult to distinguish without the positions of nodes. We briefly describe the three-step process, which generates the positional time-evolving graph: 
\begin{itemize}
    \item We sample the trajectory $\mathcal{S} = \{(x_1^{(i)}, x_2^{(i)})\}_{i=1}^{T}$ using SDE (\ref{eq:sde}), and the corresponding potential function (\ref{eq:limon_potential}) or (\ref{eq:double_potential}). 
    \item Then we choose the number of nodes \textit{n}, and assign random coordinates $(a_j, b_j), j=1, \dots, n$ to each of these nodes. The reason for that we need to know the locations of discriminating features of metastable states.
    \item In the final step, we define discriminating topological features for each state. Let $G_0$ be a complete graph. In case of the \textit{s}-well potential, we generate $G_t$, $\forall t, t = 1, \dots, T$ by drawing a circle with the centre at $(x_1^t, x_2^t)$ and the radius \textit{r} and randomly removing edges between nodes that are inside the current circle. In addition, to address the noise in the real-world data, we also remove edges outside the current circle. For double-well potential, we remove edges between nodes, which satisfy $b_j > \frac{1}{T} \sum_{i=1}^T x_1^i$.
\end{itemize}

We can see that the graph features of different states are difficult to distinguish, and the model will fail to discriminate between metastable states in the time-evolving graph. As an illustration of that, we provide Figure \ref{fig:FIG2}. There are two states \textit{A} and \textit{B} of the time-evolving graph $\mathbb{G}$ that are characterized by removed edges in the right part of $\mathbb{G}$ for the state A and in the left part of the graph for state B. Topologically, states have the same neighbourhood structures, which will result in the same points in the low-dimensional space. The same occurs for our synthetic dataset: nodes in the circles determine metastable states, and neighborhoods of these nodes are almost identical for the model.

\paragraph{\textbf{Non-positional data.}} This type of data has a dissimilar topological pattern to the positional data. The time-evolving graph is generated in the same way as the positional synthetic dataset, except instead of removing a random number of edges between nodes that fall in the circle, we remove edges between nodes in the circle in such a way that each node has the particular number of neighbours. We define the number of removed neighbors of nodes arbitrary and different for each state. 

\begin{figure}[ht]
	\centering
		\includegraphics[scale=.65]{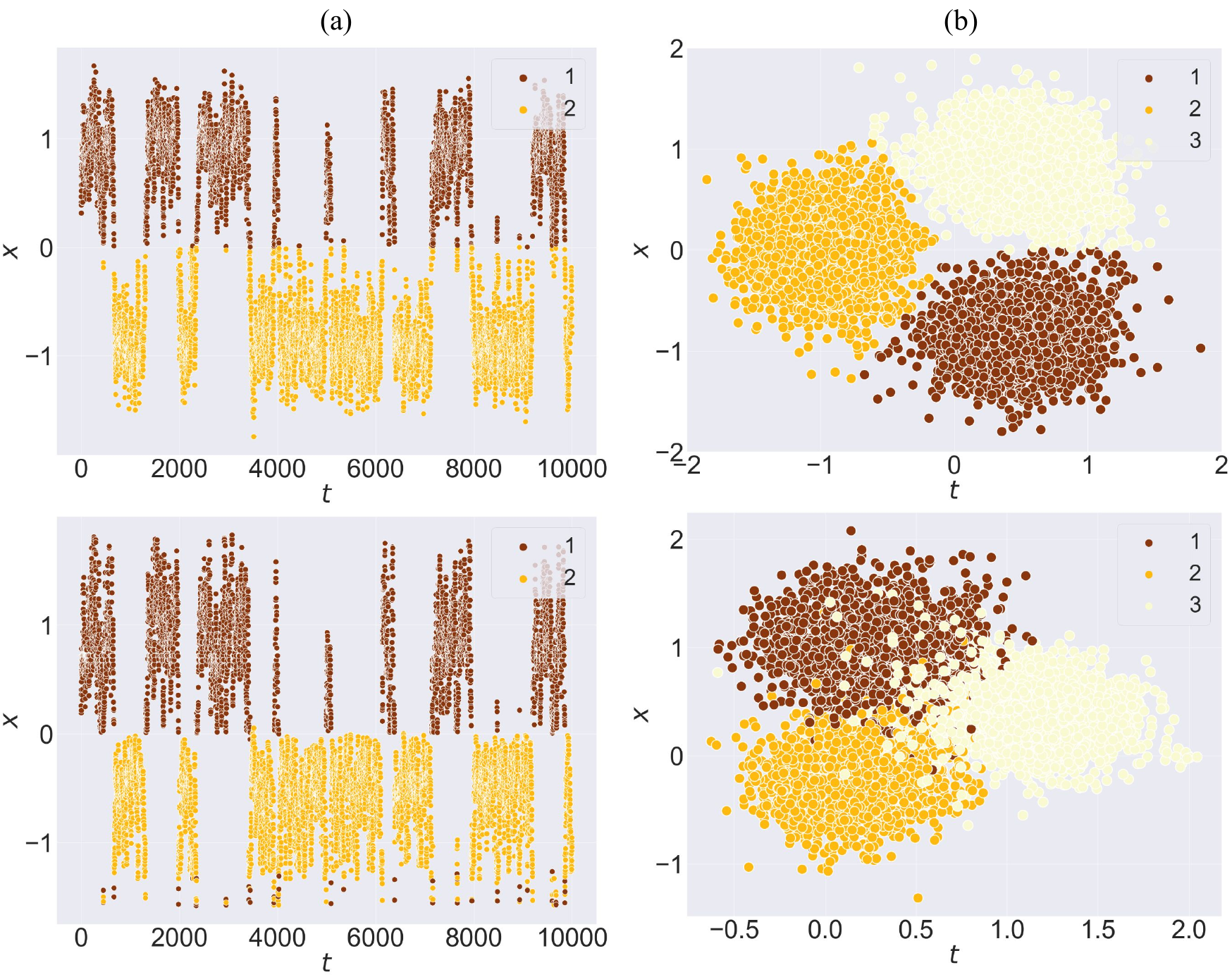}
	\caption{The comparison of an initial trajectory sampled from the SDE (\ref{eq:sde}) (\textbf{top}) and the final graph embedding for (\textbf{bottom}): (a) the \textit{npos\_2WellGraph} dataset and (b) for the \textit{pos\_3WellGraph} dataset.}
	\label{fig:FIG4}
\end{figure}

\begin{table}[ht]
  \caption{Statistics of each dataset used in this paper.}
  \begin{tabular}{|l|l|l|l|l|}
  \hline
    Name & \# Nodes & \# Edges (avg.) & \# Time steps & \# States \\
    \hline
    npos\_2WellGraph & 150 & 10821 & 10000 & 2 \\
    \hline
    pos\_2WellGraph & 100 & 4109 & 10000 & 2 \\
    \hline
    pos\_3WellGraph & 150 & 10869 & 10000 & 3\\
    \hline
    CholeraInf & 96 & 106 & 34 & 2 \\
    \hline
    MovingPic & 919 & 10602 & 658 & 2  \\
    \hline
  \end{tabular}
  \label{tab:data_stat}
\end{table}

\subsubsection*{Real-world dataset}
\paragraph{\textbf{MovingPic.}} This dataset, originally introduced in \cite{MovingPicture}, is the first real-world dataset on which we evaluate our model. In this study, one male and one female  were sampled daily at three body sites (gut, skin, and mouth) for 15 months and for 6 months, respectively. To obtain a time-evolving graph, we pre-process Operational Taxonomic Units (OTU) that contains the number of 16S rDNA marker gene sequences that are observed for each taxonomic unit in each sample. Let $D \in \mathbb{R}^{T \times p}$ be an OTU table, where \textit{T} is the number of time points and \textit{p} is the number of OTUs. As this data does not have any obvious perturbations, such as antibiotics exposure or diseases, which could potentially create a metastable structure, an artificial noisy signal is added to the data. The Pearson correlation between two OTUs is computed, and then the initial time-snapshot graph is constructed. To construct time-snapshot graphs at each time step, the authors of \cite{graphkke} use the OTU table to remove edges between nodes. If the OTU count for a particular node is zero, then the edge is removed between this node and its neighboring nodes. 


\paragraph{\textbf{CholeraInf.}} This dataset has been introduced in a study about the recovery from Vibrio Cholera infection \cite{CholeraInfOriginal}. Here, faecal microbiota were collected from seven cholera patients from disease (state 1) through recovery (state 2) periods. Moreover, in our experiment, we use the microbiome of one patient, since the variation in the microbiome constitution among individuals can have an impact on the result of the model. The time-evolving graph is obtained in the same way as it has been done for the MovingPic dataset.

\subsection*{Visualization and comparative analysis}
In this part, we focus on verifying the qualitative performance of our model. As a first experiment, we visualize the resulting graph embedding to evaluate how separated the metastable states are in the low-dimensional space. As a second experiment, we compare our model with the following methods: a simple baseline chosen from state-of-the-art methods for dimensional reduction, namely, Principal Component Analysis (PCA), two kernel-based methods graphKKE \cite{graphkke} and WL kernel \cite{WL}, and two graph representation learning methods node2vec \cite{node2vec}, and graph2vec \cite{graph2vec}. 

\begin{itemize}
    \item Principal Component Analysis (PCA) is a method for dimensional reduction. To be able to apply this method to the time-evolving graph, we flatten an adjacency matrix of each time-snapshot graph into a vector.
    
    \item The graphKKE approach is proposed for learning the embedding of a time-evolving graph with metastability. It is a graph kernel-based method that combines a transfer operator theory and a graph kernel technique. 

    \item The WL kernel decomposes graphs into rooted subgraphs using a relabelling process and computes feature vectors based on the number of initial and updated labels. 

    \item The graph2vec approach projects the set of static graphs, and it comprises two main components: 1) Weisfeiler-Lehman relabeling process and 2) the Skip-gram procedure from doc2vec \cite{doc2vec}.
    
    \item The node2vec algorithm is a node representation method that uses breadth-first search and depth-first search to extract local and global information from the static graph.
    
\end{itemize}

\paragraph{\textbf{Evaluation metric. }} In order to conduct the comparison analysis, we use a standard clustering evaluation metric --- Adjusted Rand Index(ARI). The ARI values lie in the range $[-1;1]$ with 0 representing random clustering and 1 being the highest correspondence to the ground-truth data. 

\paragraph{Experimental setup.} First, we examine the evolution of the graph embedding by visualizing it at the beginning, in the middle and at the end of the training of the model. To do so, we use the graph embedding $\hat{g} =\{\hat{g}_1, \dots, \hat{g}_{T}\}$, where $\hat{g}_i \in \mathbb{R}^{d}$ with $d=2$. For all synthetic datasets, we set the hyperparameter \textit{l} to be 3, the batch size to be 64, and the number of epochs to be 200 for both pos\_3WellGraph and npos\_2WellGraph. For real-world data, the batch size is set to be 64 for MovingPic and 6 for the CholeraInf dataset. We use the Adam optimizer with default parameters, the number of heads of the Transformer 4 and the number of layers in the Transformer 3. As the graphKKE method approximates the eigenfunctions of a transfer operator, the dimension of the graph embedding equals the number of metastable states in the time-evolving graph. This means that we need to apply a dimensional reduction method to be able to visualize it. Thus, PCA is applied to the output of graphKKE with the number of components to be 2. We also apply PCA to the flattened adjacency matrices with the number of components to be 2. Moreover, since we are interested in whether metastable states of the original space correspond to the clusters of points in the reduced space, the points of the graph embeddings are colored according to original ground truth metastable states.

\begin{figure}[ht]
	\centering
		\includegraphics[scale=.65]{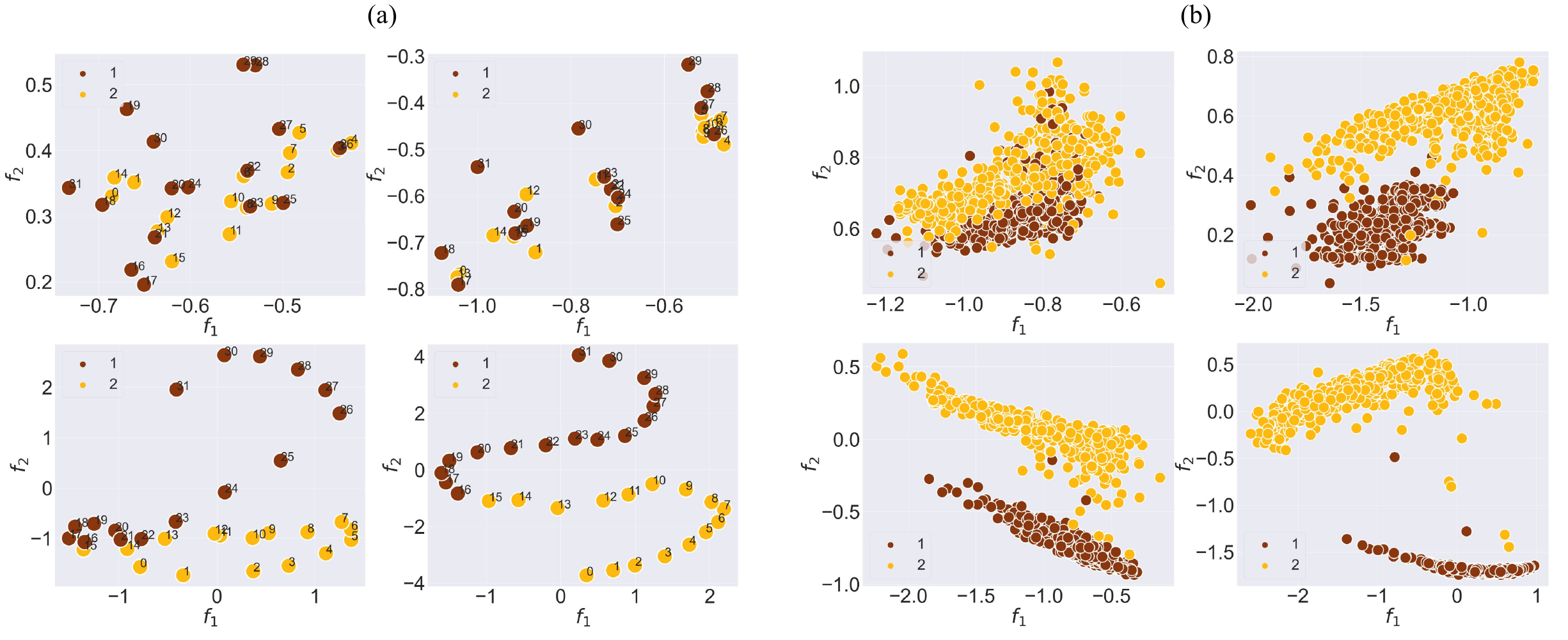}
	\caption{The evolution of the graph embedding of the time-evolving graph $\mathbb{G}$ during the training of our model on (a) CholeraInf and (b) MovingPic. The points are colored according to ground-truth labels.}
	\label{fig:FIG5}
\end{figure}

For the comparison analysis we obtain graph embeddings from other methods in the following way. We set a number of dimensions of graph embeddings to be 32 for our method, node2vec, graph2vec and PCA. We use the implementations for node2vec and graph2vec with the default hyperparameters provided by the authors. In the graphKKE method, the number of dimensions of the final graph embedding equals to the number of metastable states in the time-evolving graph. Finally, we apply the \textit{k}-means method to cluster points of the final graph embeddings of each method. However, node2vec is developed to learn node representations, that is why, to obtain embeddings of entire time-snapshots graph, we average node embddings of each time-snapshot graph.

\begin{figure}[ht]
	\centering
		\includegraphics[scale=.6]{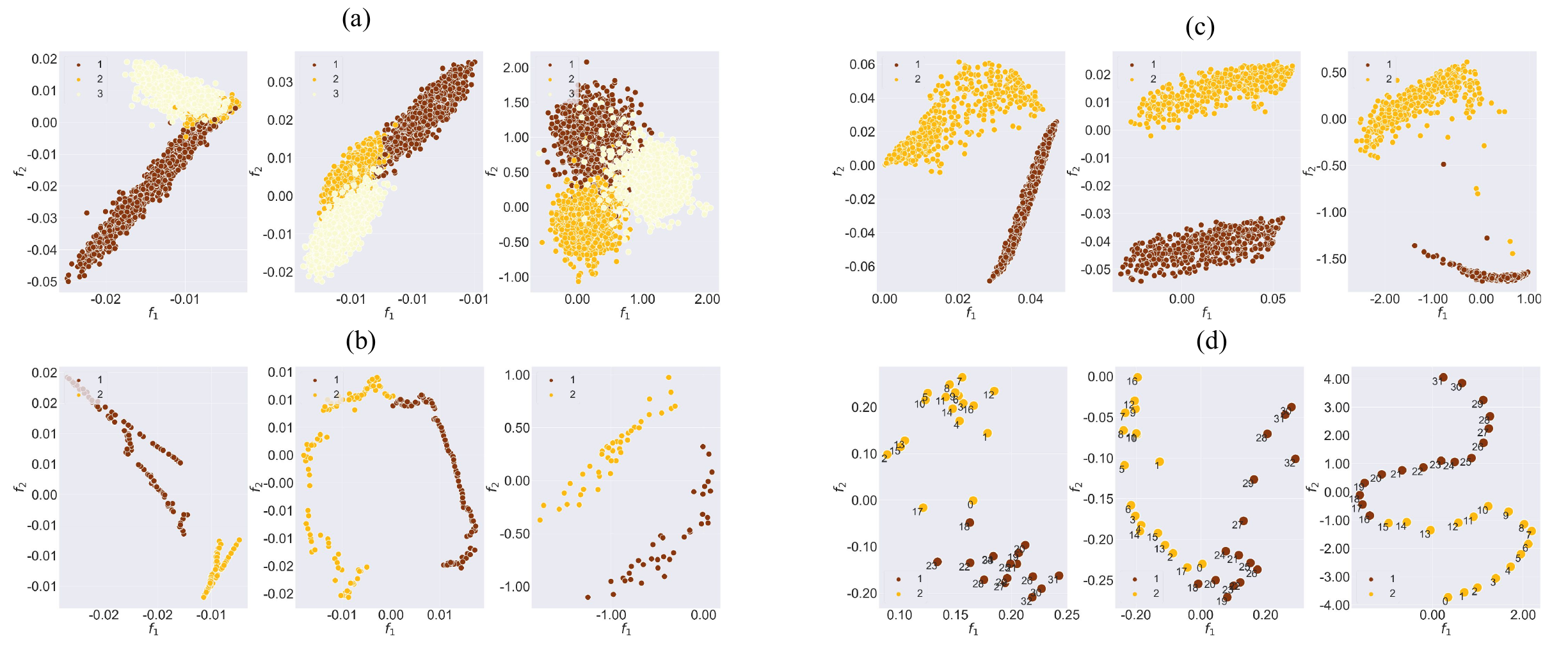}
	\caption{The graph embeddings of the time-evolving graph $\mathbb{G}$ for (a) npos\_2WellGraph, (b) pos\_3WellGraph, (c) MovingPic and (d) CholeraInf. From left to right: PCA on adjacency matrices, PCA on eigenfunctions of graphKKE and the result of our model.}
	\label{fig:FIG6}
\end{figure}

\begin{table}[ht]
  \caption{Adjusted Rand Index (ARI) for the comparative analysis on the graph clustering task. ARI close to 1 corresponds to greater accuracy in correctly identifying the ground truth states, and an ARI value close to 0 stands for random clustering.}
  \begin{tabular}{|l|l|l|l|l|l|}
  \hline
    Dataset & graphKKE & WL kernel & graph2vec & node2vec & our model \\
    \hline
    npos\_2WellGraph & 0.99  & 0.98 & 0.963 & 0.000 & 0.99 \\
    \hline
    pos\_2WellGraph & 0.97 & 0.97 & 0.046 & 0.000 & 0.82\\
    \hline
    pos\_3WellGraph & 0.93 & 0.11  & 0.003 & 0.000 & 0.80\\
    \hline
    MovingPic & 0.99 & 0.56 & 0.42 & 0.078 & 0.99  \\
    \hline
    CholeraInf & 0.88 & 0.66 & 0.29 & -0.017 & 0.87 \\
    \hline
  \end{tabular}
  \label{tab:baseline}
\end{table}

\paragraph{\textbf{Result and Discussion: synthetic data.}} 
The evolution of the graph embedding during the training for both synthetic datasets --- npos\_2WellGraph and pos\_3WellGraph --- are illustrated in Figure \ref{fig:FIG3}. The visualization demonstrates that during the training, our model tends to capture the underlying metastable structure in the time-evolving graph. Moreover, at the end of the training, we see that our model learns the graph embedding maintaining the initial metastable dynamics. In the case of the npos\_2WellGraph dataset, there is no obvious split between the two metastable states, the reason is that the initial SDE trajectory has points that are located on the boundary between two metastable states. Furthermore, we compare the initial SDE trajectory and the final graph embedding obtained from our model with $d=1$ for npos\_2WellGraph and with $d=2$ for pos\_3WellGraph. The result for npos\_2WellGraph is presented in Figure \ref{fig:FIG4}(a) which shows that two trajectories are almost identical. The same result can be seen for pos\_3WellGraph in Figure \ref{fig:FIG4}(b). These results indicate that the model is capable of extracting the underlying metastable dynamics in the time-evolving graph. 

In Table \ref{tab:baseline}, we can see that the results of visualization are reinforced by the high ARI values of our model. From the table can also be seen that the graphKKE method outperforms our model in case of the pos\_2WellGraph and pos\_3WellGraph datasets. However, if we aim to have lower dimensionality of the graph embedding, then this method will fail to produce the same clustering accuracy. As the evidence the visualization of the graph embedding obtained with graphKKE (Figure \ref{fig:FIG6}), we see that graphKKE+PCA fails to produce a visualization with clear separated metastable states. Moreover, considering the results of other graph representation learning (Table \ref{tab:baseline}), node2vec fails completely to learn the graph embedding of the time-evolving graph and graph2vec performs poorly on all synthetic datasets except npos\_2WellGraph. It remains unclear whether 
graph2vec struggles to identify states in the positional data because states do not have unique topological patterns, or because this method is not meant to capture temporal changes.

\paragraph{\textbf{Result and Discussion: real-world data.}} In the case of real-world datasets, the evolution of the graph embedding during the training for MovingPic and CholeraInf are presented in Figure \ref{fig:FIG5}. As it was in the case of synthetic datasets, our method is also able to identify the metastable behaviour in the time-evolving graph and preserve it in the new space. For CholeraInf we have added time points from the original dataset to see if the new space has the same time order as it was in the original high-dimensional space. If we compare the visualization of graph embedding from other methods, the result for MovingPic shown in Figure \ref{fig:FIG6}(c) shows that all methods give relatively the same visualization. However, for the CholeraInf (Figure \ref{fig:FIG6}(d)) our model preserves consecutive time points in the new space which indicates that one metastable state (healthy) follows another metastable states (ill).

The second part of this experiment aims at comparing our model with other dimensional reduction methods in the clustering task. Again from the Table \ref{tab:baseline} it is evident that our model performs significantly better that WL kernel, graph2vec and node2vec. Node2vec performs poorly across all datasets, which is the result of a lower order substructure embedding method meaning that it can model only local similarities and fails learn global topological similarities.

\begin{figure}[ht] 
	\centering
	\includegraphics[scale=.65]{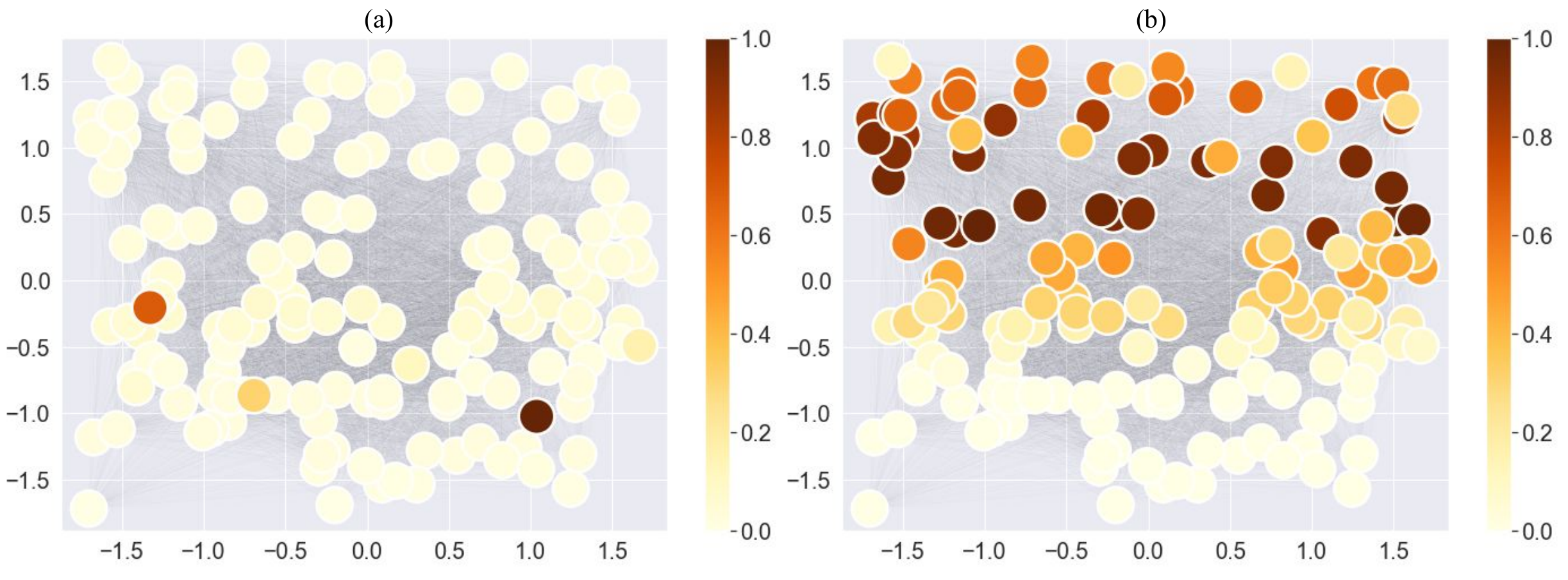}
	\caption{Fully-connected graphs for npos\_2WellGraph dataset with nodes colored based on the relevance score of the LRP method, which are summed across 2 states: (a) State 1 (b) State 2. The locations of 2 states are obtained by clustering points of the graph embedding of the time-evolving graph via \textit{k}-means.}
	\label{fig:FIG7}
\end{figure}

\subsection*{Interpretebility}
\label{sec: explainability}
An improved understanding of how the microbiome contributes to health and well-being can drive and accelerate the development of microbiome-based treatments. The most important question, which has not been answered yet, is which species or interactions of species are responsible for or affected by the changes which microbiome undergoes from one state (healthy) to another state (diseased or antibiotic exposure). The presence of such valuable information can significantly improve modern treatments of various diseases. We assume that if it is feasible for the model to successfully find and discriminate metastable states, then there might be topological features in the time-evolving graph that make these metastable states different. Therefore, the main objective of this section is to provide an insight into to which extent the model learns metastable states based on true discriminating topological features. And with regard to real-world data, we aim to find topological features of the time-evolving graph that make the two states, cholera infection period and recovery period, different.

To achieve this, we will use an approach\cite{TransformerIB} which is based on layer-wise relevance propagation (LRP) \cite{LRP}. LRP is the family of explanation methods that leverages the layered structure of the network. It explains the prediction of a neural network classifier by backpropagating the neuron activation on the output layer to the previous layers until the input layer is reached. 

The authors of the paper \cite{TransformerIB} address the lack of the conservation property in the attention mechanism, which is an essential assumption of LRP and the numerical issues of the skip connections by applying a normalization to the computed relevance score. Moreover, they make use of the attention weights and propose to compute the final relevance scores by multiplying the relevance score of each Transformer layer with the gradient of the corresponding attention matrix summed up across the ``head`` dimension.  

Unlike original LRP and the approach mentioned in the last paragraph, where the decomposition starts from the classifier output corresponding to the target class, we have a similarity model that rather measures how similar graph embeddings of the time-snapshot graphs $G_t$ and $G_{t+1}$ are. For this reason, we start the redistribution from the layer where we compute the graph embedding $\mathbf{\hat{g}}_t$ until the input layer is reached, and the final relevance is computed. We compute a relevance score for each time-snapshot graph in the test set. To obtain discriminating features of the whole state, we sum up relevance scores of time-snapshot graphs of each state.

\paragraph{\textbf{Result and Discussion: synthetic dataset.}}
We conduct this experiment on the npos\_2WellGraph and the CholeraInf datasets. The result for npos\_2WellGraph is demonstrated in Figure \ref{fig:FIG7}. From the result it is clear that the model can find topological features in the time-evolving graph that are unique for each metastable state. However, the interpretation of state 1 (Figure \ref{fig:FIG7}b) highlights all nodes in the upper part of the time-snapshot graph, which is a true discriminating feature, wheres the interpretation of state 0 in turn shows only 4 nodes in the lower part of the time-snapshot graph. 

There is a necessity to mention that we have modelled the synthetic datasets in such a way that we know the location of nodes in the time-snapshot graphs. In case of real-world datasets, we do not have coordinates of nodes. 

\paragraph{\textbf{Result and Discussion: real-world dataset.}}
We have assessed the interpretation of our model on the synthetic dataset, and now we focus on obtaining relevance scores for the real-world dataset, namely, CholeraInf. Unlike the synthetic dataset, we do not know the ground truth discriminating features for this dataset. Moreover, to visualize the interpretation, we use a correlation matrix that has been computed based on the OTU table (see more details about how this data has been pre-processed in \cite{graphkke}).

\begin{figure}[ht] 
	\centering
		\includegraphics[scale=.55]{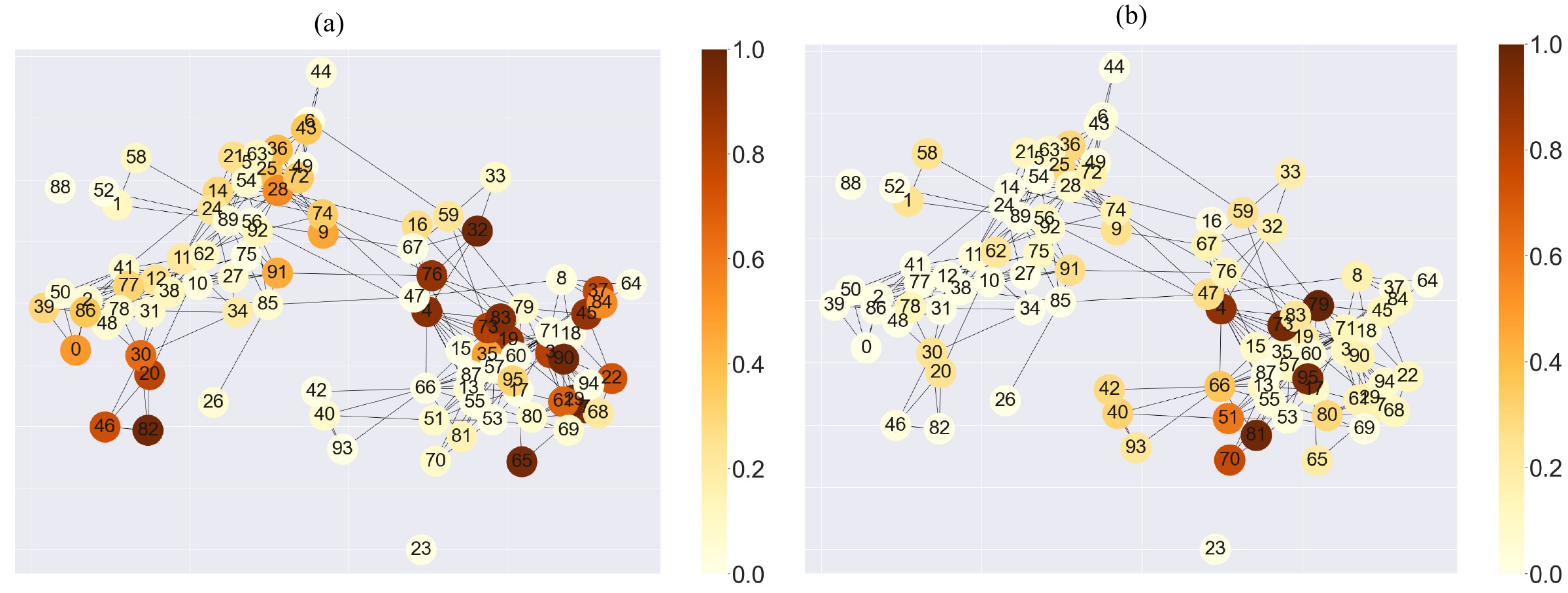}
	\caption{Co-occurrence interaction graphs of CholeraInf dataset with nodes colored based on relevance scores of the LRP method, which are summed across each state: (a) diarrhea period and (b) recovery period. The locations of states are obtained by clustering points of the graph embedding of the time-evolving graph via \textit{k}-means. The dark brown colour indicates nodes with the highest importance.}
	\label{fig:FIG8}
\end{figure}

The result of LRP is presented in Figure \ref{fig:FIG8}a for the diarrhea period (state 1) and in Figure \ref{fig:FIG8}b for the recovery period (state 2). We can see that we have two different sets of nodes with high relevance score, yet there are the same nodes in both states that are significant for the model. For example, the relevance scores of node 73 and node 4 are high in both states.

Further study of these results is needed to investigate if these interpretations have biological meaning. For instance, there have been done numerous works \cite{Langdon2016TheEO} that are mainly focused on statistical analysis to justify which bacteria/species are affected by, or on the contrary, cause shifts in microbiome compositions. Using the detected species from these works, we can compare them with the nodes that have shown the biggest impact on the model output according to both the LRP approach and the visualization of attention weights.

\section*{Discussion}

We have presented a new approach that can simplify the analysis of time-evolving graphs with assumed metastability. Through an extensive set of experiments on both synthetic and real-world datasets, we have demonstrated that our approach is capable of projecting a time-evolving graph into a low-dimensional space retaining metastable properties of the system. Moreover, we have illustrated one of the possible applications of this approach to microbiome data that enhances the analysis of metagenomic data in a way that it takes into account a huge number of interactions among species. We have shown that through explaining the output of the model, we can find topological graph features, such as nodes or edges, that make the model arrive at a certain graph embedding. Concerning microbiome data, it means that our method coupled with a proper interpretation strategy can help to reveal underlying disease patterns in the data.

There are several directions for future work: 1) how to construct a time-evolving graph from metagonomic data such that it contains a real dynamics occurring in the microbiome; 2) further biological analysis of results obtained from the interpretability of the model; 3) visualization of topological graph features, such as nodes and edges, that have impacted the model the most, and 4) mathematical explanation of how the model learns a graph embedding of the time-evolving graph maintaining metastable dynamics.

\section*{Method}
\label{sec:proposed_method}
We first briefly introduce all necessary notations and definitions, which are used in the paper, and state the problem. 

\paragraph{\textbf{Definitions.}}A graph $G$ is a pair $(V, E)$ with a non-empty set of nodes $V(G)$ and a set of edges $E(G) = \{(v_i, v_j) \mid v_i, v_j \in V\}$. The set $V(G)$ often represents the objects in the data and $E(G)$ the relations between objects. We define \textit{the adjacency matrix} of the graph $G$ as the $n \times n$ matrix $A$ with $A^{ij} = 1$ if the edge $(v_i, v_j) \in E(G)$, and 0 otherwise, where $n = \mid V \mid$.

Next, we define a metastability property, which was first mentioned in \cite{graphkke}. Consider a time-evolving graph $\mathbb{G}$ as a sequence of graphs $\mathbb{G} = (G_1,\dots, G_{T})$ at consecutive time points $\{1,\dots, T\}$ for some $T \in \mathbb{N}$, and $G_t$ being a time-snapshot of $\mathbb{G}$ at time $t$. The time-evolving graph $\mathbb{G}$ exhibits {\em metastable} behavior if $\mathbb{G}$ can be partitioned into $s$ subsets $\mathbb{G} = \mathbb{G}_1 \cup \dots \cup \mathbb{G}_{s}$ for some $s \ll T$ such that for each time point $t \in \{1,\dots, T\}$ and $i, j = 1, \dots, s$, we have the following:
\begin{equation}
\label{eqn:metastability}
\begin{cases}
    P(G_{t + 1} \in \mathbb{G}_i \mid G_t \in \mathbb{G}_j) \ll 1, \text{ if } i \neq j\\
    
    P(G_{t + 1} \in \mathbb{G}_i \mid G_t \in \mathbb{G}_j) \approx 1, \text{ if } i = j,
\end{cases}
\end{equation}
where $P(\cdot)$ is a transition probability, and $\mathbb{G}_1, \dots, \mathbb{G}_{s}$ are called metastable states of the time-evolving graph $\mathbb{G}$, where $s$ is the number of states.

\paragraph{\textbf{Problem statement.}} We define our problem as follows: \textit{Given a time-evolving graph $\mathbb{G} = (G_1, \dots, G_{T})$ with assumed metastability property (\ref{eqn:metastability}), we aim to represent each time-snapshot $G_t$ as a vector in a low-dimensional space $\mathbb{R}^d$, maintaining the metastable behaviour of $\mathbb{G}$, where $d$ is a number of dimensions of the reduced space.}

In this section, we describe how we train the model to embed time-snapshot graphs into a low-dimensional space, maintaining the metastable behaviour of the graph. We first use the Transformer  \cite{AttentionWhatYouNeed} to compute the embedding of a time-snapshot graph. Further, we add a recurrent mechanism to the Transformer that facilitates the learning of temporal changes across consecutive time-snapshot graphs. Finally, we use contrastive learning to make representations of consecutive time-snapshots graphs, which share metastable behaviour, close.

\begin{algorithm}
\caption{Main learning algorithm} \label{alg:main_model}
\KwIn{\\ \textit{l}: the number of historical representations;\\
$\mathbb{G} = \{G_1, \dots, G_{T}\}$: a time-evolving graph such that each time-snapshot graph $G_t = (x_t, A_t)$;\\
$\mathbf{z}_{init}$: a randomly initialized learnable master node;
R: the Transformer; \\
$W, W^{(1)}, W^{(2)}, b$: parameters of the model;
}
$\mathcal{L} = 0$; \\
\vspace{0.2cm}
\For{\textit{sampled mini batch of indices} $\{t_k\}_{k=1}^N$}{
$\mathbf{z}_{curr} = \mathbf{z}_{init}$; $\mathbf{z}_{next} = \mathbf{z}_{init}$ \\
\vspace{0.1cm}

\For{$j = 1$ \KwTo $l$}{
    \vspace{0.2cm}
    \CommentSty{\textcolor{gray}{\#time-snapshot graphs at time \textit{t}}} \\
    Select time-snapshot graphs $\{G_{t_k + j}\}_{k=1}^N$ \\
    
    $\mathbf{z}_{curr} = R(\mathbf{z}_{curr}, x_{t_k + j}, A_{t_k + j})$ \\
    
    \vspace{0.2cm}
    \CommentSty{\textcolor{gray}{\#time-snapshot graphs at time  \textit{t+1}}} \\
    
    Select time-snapshot graphs $\{G_{(t_k+1) + j}\}_{k=1}^N$ \\
    
    $\mathbf{z}_{next} = \mathcal{R}(\mathbf{z}_{next}, x_{(t_k + 1) + j}, A_{(t_k + 1) + j})$ \\                 
    
    \vspace{0.2cm}
    \CommentSty{\textcolor{gray}{\#final graph embeddings}} \\
    $\mathbf{\hat{g}}_{curr} = W \mathbf{z}_{curr} + b$ \\ 
    $\mathbf{\hat{g}}_{next} = W \mathbf{z}_{next} + b$
    
    \vspace{0.2cm}
    \CommentSty{\textcolor{gray}{\#projection for contrastive learning}} \\
    $\mathbf{g}_{curr}$ = $W^{(2)} \sigma(W^{(1)} \mathbf{\hat{g}}_{curr})$ \\  
    $\mathbf{g}_{next}$ = $W^{(2)} \sigma(W^{(1)} \mathbf{\hat{g}}_{next})$
    
    \vspace{0.2cm}
    \CommentSty{\textcolor{gray}{\#pairwise similarity}} \\
    $sim = \frac{\mathbf{g}_{curr}^T \cdot \mathbf{g}_{next}}{\parallel \mathbf{g}_{curr} \parallel \cdot \parallel \mathbf{g}_{next} \parallel}$

    $\mathcal{L} = \mathcal{L} + \sum_{i=1}^{N} -\log \Big( \frac{\exp{\big(sim_{i, i} /
    \tau \big)}}{\sum_{j=0}^N \exp{\big(sim_{i, j} / \tau \big)}} \Big)$ \\
}
}
\Return $\mathcal{L}$
\end{algorithm}

\paragraph{\textbf{Transformer.}}The Transformer is currently the state-of-the-art method in the field of NLP, where it has shown tremendous success for handling long-term sequential data. Recently, it has become a leading tool in other domains such as computer vision \cite{ViT2020} and graph representation learning \cite{GraphAttentionNetwork, GeneralizationOfTransformer}. We use the encoder part of the Transformer to learn node embeddings in each time-snapshot graph. The encoder has several stacked multi-head attention layers followed by a feed-forward layer. There is a residual connection around each of these two sub-layers that is also followed by a normalization layer.

Intuitively, the self-attention in time-snapshot graphs relates different nodes of the graph in order to compute a new representation of every node in the graph to which we refer as a node embedding.

\paragraph{\textbf{Input.}}Let $\mathbb{G} = \{G_1, ..., G_{T}\}$ be a time-evolving graph with node features $\{x_t^v\}_{v \in V(G_t)}$, $t = 1, \dots, T$. The input node features of each time-snapshot graph $G_t$ are embedded to $\textit{d}_m$-dimensional latent features via a linear projection and added to pre-computed node positional encodings. We demonstrate on two synthetic datasets with different topological structures that our model performs well in both cases: with positional information of nodes and without it. 
Moreover, in order to capture the topological structure of a single time-snapshot graph $G_t$, we feed an adjacency matrix $A_t$ to the Transformer as a mask, and we set attention weights to 0 whenever the corresponding adjacency matrix entries are 0. 

\paragraph{\textbf{Model architecture.}} Further, we explain important details of the training of the model. The overview of the model architecture can be found in Figure \ref{fig:FIG1}.

Let $\mathcal{T} = \{t_k\}_{k=1}^{N}$ be a set of randomly sampled time points, and $\mathbb{G}_{\mathcal{T}} = \{G_{t_1}, \dots, G_{t_{N}}\}$ be a mini-batch of time-snapshots graphs sampled from $\mathbb{G}$. To facilitate the learning of temporal changes, we share the embedding of a time-snapshot graph with the consecutive time-snapshot graph in the temporal sequence. We define a master node that is connected to all nodes in the time-snapshot graph. Initially, the master node is represented as a learnable, randomly initialized vector. The Transformer computes the embedding of the master node which we consider as a graph embedding. This graph embedding is then passed as the initial master node to the consecutive time-snapshot graph in the temporal sequence. We control the length of temporal sequence with the hyperparameter l. Moreover, since we connect the master node with all other nodes in each time-snapshot graph, the size of the adjacency matrix changes, $A_t \in \mathbb{R^{(n+1) \times (n+1)}}$.

Formally, we update the graph embedding $z_t$ of the time-snapshot graph $G_t$ recursively as follows:
\begin{equation*}
    z_t = R(z_{t-1}, x_{t}, A_{t}),
\end{equation*}
where $R$ is the Transformer that updates node embeddings discussed, $z_{t} \in \mathbb{R}^{ d_m}$ is a master node, $x_{t}$ is a vector of node features, and  $A_{t} \in \mathbb{R}^{(n+1) \times (n+1)}$ is an adjacency matrix of $G_{t}$. 

Finally, we project the graph embedding $z_{t} \in \mathbb{R}^{d_m}$ of the time-snapshot graph $G_{t}$ into the space with the dimension $d$, where a downstream task is defined. We denote the final embedding of the time-snapshot graph with $\hat{g}_{t} \in \mathbb{R}^d$:
\begin{equation*}
    \hat{g}_{t} = W z_{t}  + b
    \label{eqn:task_projection}
\end{equation*}
with $W$ and $b$ being learnable parameters. 

We use two hidden layers and a non-linear activation function in order to project the graph embedding $\hat{g}_{t}$ into the space, where the contrastive learning is defined, as it is done in \cite{SimpleContrastiveLearning}.

Furthermore, we explain how we use the contrastive learning to make embeddings of consecutive time-snapshots graphs preserve the metastable behaviour in the low-dimensional space.

\paragraph{\textbf{Contrastive learning.}} Intuitively, contrastive representation learning can be considered as learning by comparing. Thus, the goal is to find a low-dimensional space where samples from the same instance are pulled closer and samples from different instances are pushed apart. Formally, given a vector of input samples $x_i, i=1, \dots, N$ with a corresponding labels $y_i \in \{1, \dots, C\}$ among \textit{C} classes, contrastive learning aims to learn a function $f_\theta(x)$ that can find the low-dimensional representation of $x$ such that examples from the same class have similar representations and samples from different classes are far away from each other in a new space. One always needs to have negative and positive samples to apply contrastive learning, For this reason, we make the following assumption.

\textit{Assumption.} According to the definition of metastability (\ref{eqn:metastability}), the probability of two consecutive time-snapshot graphs $G_{t}$ and $G_{t+1}$ being similar is almost 1 and so should be the probability for their graph embeddings $\hat{g}_{t}$ and $\hat{g}_{t+1}$. 

In other words, we consider a pair of graph embeddings $(\hat{g}_{t}, \hat{g}_{t+1})$ as a positive pair and pairs $(\hat{g}_{t}, \hat{g}_{t+\tau}), t+\tau \in \{2, \dots, N\}$ as negative pairs, where $\tau$ is randomly sampled. It is feasible for negative samples to be of the same metastable states, but at different time points. 

Given graph embeddings $g_{t}$ and $g_{t+1}$ for $G_{t}$ and $G_{t + 1}$ as the output of our model, we first compute the similarity between $g_{t}$ and $g_{t+1}$:
\begin{equation*}
    sim_{t, t+1} = \frac{g_{t}^T \cdot g_{t+1}}{\parallel g_{t} \parallel \cdot \parallel g_{t+1} \parallel}.
\end{equation*}

With the above assumption, diagonal elements of \textit{sim} represent positive pairs and off-diagonal elements negative pairs. Then, the similarity score is fed to the Noise Contrastive Estimator (NCE) loss: 
\begin{equation*}
    \mathcal{L} = -\log \Big( \frac{\exp{\big(sim_{t, t+1} /
    \tau \big)}}{\sum_{j=1}^T \exp{\big(sim_{t, j} / \tau \big)}} \Big).
\end{equation*}

Minimizing this loss function forces the parameters of the model to be tuned such that graph embeddings of two consecutive time-snapshot graphs is as close as possible. 

\paragraph{\textbf{Positional Encoding}}
Most graph neural networks learn structural node information that is invariant to the node positions. However, there are cases when topological information is not enough. To demonstrate this, we conduct experiments on two different synthetic datasets. The first data has metastable states with defined graph features that can be distinguished only with the position information of nodes. Each metastable state in the second data has specific graph features, which are easily distinguished with just topological information. 

To incorporate the positional information, we use the same positional encoding as in \cite{gehring17a}:
\begin{equation*}
    p_{pos, 2i} = \sin{pos/10000^{2i/d_m}}
\end{equation*}
\begin{equation*}
    p_{pos, 2i+1} = \cos{pos/10000^{2i/d_m}},
\end{equation*}
where \textit{pos}, \textit{i} and $d_m$ denote a position of the node in the time-snapshot graph, the dimension in the positional encoding and the dimension of node embedding, respectively.

Through a set of various experiments in the next section, we demonstrate on synthetic and real-world datasets that our method is capable of learning a graph embedding of the time-evolving graph.


\bibliography{main}

\begin{thebibliography}{10}
\urlstyle{rm}
\expandafter\ifx\csname url\endcsname\relax
  \def\url#1{\texttt{#1}}\fi
\expandafter\ifx\csname urlprefix\endcsname\relax\def\urlprefix{URL }\fi
\expandafter\ifx\csname doiprefix\endcsname\relax\def\doiprefix{DOI: }\fi
\providecommand{\bibinfo}[2]{#2}
\providecommand{\eprint}[2][]{\url{#2}}

\bibitem{InflammatoryMicrobiome}
\bibinfo{author}{Joossens, M.} \emph{et~al.}
\newblock \bibinfo{journal}{\bibinfo{title}{Dysbiosis of the faecal microbiota
  in patients with crohn's disease and their unaffected relatives}}.
\newblock {\emph{\JournalTitle{Gut}}} \textbf{\bibinfo{volume}{60}},
  \doiprefix\url{10.1136/gut.2010.223263} (\bibinfo{year}{2011}).

\bibitem{InflammarotyMicrobiome1}
\bibinfo{author}{Mottawea, W.} \emph{et~al.}
\newblock \bibinfo{journal}{\bibinfo{title}{Altered intestinal
  microbiota–host mitochondria crosstalk in new onset crohn's disease}}.
\newblock {\emph{\JournalTitle{Nature Communications}}}
  \textbf{\bibinfo{volume}{7}} (\bibinfo{year}{2016}).

\bibitem{ObesityMicrobiome2}
\bibinfo{author}{Menni, C.} \emph{et~al.}
\newblock \bibinfo{journal}{\bibinfo{title}{Gut microbiome diversity and
  high-fibre intake are related to lower long-term weight gain}}.
\newblock {\emph{\JournalTitle{International Journal of Obesity}}}
  \textbf{\bibinfo{volume}{41}}, \bibinfo{pages}{1099--1105},
  \doiprefix\url{10.1038/ijo.2017.66} (\bibinfo{year}{2017}).

\bibitem{CancerMicrobiome1}
\bibinfo{author}{Sánchez-Alcoholado, L.} \emph{et~al.}
\newblock \bibinfo{journal}{\bibinfo{title}{The role of the gut microbiome in
  colorectal cancer development and therapy response}}.
\newblock {\emph{\JournalTitle{Cancers}}} \textbf{\bibinfo{volume}{12}},
  \doiprefix\url{10.3390/cancers12061406} (\bibinfo{year}{2020}).

\bibitem{CancerMicrobiome2}
\bibinfo{author}{Parida, S.} \& \bibinfo{author}{Sharma, D.}
\newblock \bibinfo{journal}{\bibinfo{title}{The microbiome and cancer: Creating
  friendly neighborhoods and removing the foes within}}.
\newblock {\emph{\JournalTitle{Cancer Research}}}
  \textbf{\bibinfo{volume}{81}}, \bibinfo{pages}{790--800},
  \doiprefix\url{10.1158/0008-5472.CAN-20-2629} (\bibinfo{year}{2021}).

\bibitem{CancerMicrobiome3}
\bibinfo{author}{Chattopadhyay, I.} \emph{et~al.}
\newblock \bibinfo{journal}{\bibinfo{title}{Exploring the role of gut
  microbiome in colon cancer}}.
\newblock {\emph{\JournalTitle{Applied Biochemistry and Biotechnology}}}
  \textbf{\bibinfo{volume}{193}}, \bibinfo{pages}{1780 -- 1799}
  (\bibinfo{year}{2021}).

\bibitem{CancerMicrobiome4}
\bibinfo{author}{Chambers, L.} \emph{et~al.}
\newblock \bibinfo{journal}{\bibinfo{title}{The microbiome and gynecologic
  cancer: Current evidence and future opportunities}}.
\newblock {\emph{\JournalTitle{Current Oncology Reports}}}
  \textbf{\bibinfo{volume}{23}} (\bibinfo{year}{2021}).

\bibitem{StatisticApproach1}
\bibinfo{author}{Mukherjee, A.} \emph{et~al.}
\newblock \bibinfo{journal}{\bibinfo{title}{Bioinformatic approaches including
  predictive metagenomic profiling reveal characteristics of bacterial response
  to petroleum hydrocarbon contamination in diverse environments.}}
\newblock {\emph{\JournalTitle{Scientific Reports}}}
  \textbf{\bibinfo{volume}{7}}, \doiprefix\url{10.1038/s41598-017-01126-3}
  (\bibinfo{year}{2017}).

\bibitem{MovingPicture}
\bibinfo{author}{Caporaso, J.} \emph{et~al.}
\newblock \bibinfo{journal}{\bibinfo{title}{Moving pictures of the human
  microbiome}}.
\newblock {\emph{\JournalTitle{Genome biology}}} \textbf{\bibinfo{volume}{12}},
  \bibinfo{pages}{R50}, \doiprefix\url{10.1186/gb-2011-12-5-r50}
  (\bibinfo{year}{2011}).

\bibitem{PhysicLandscape}
\bibinfo{author}{Chang, W.~K.} \& \bibinfo{author}{VanInsberghe, L.,
  David~andKelly}.
\newblock \bibinfo{journal}{\bibinfo{title}{Topological analysis reveals state
  transitions in human gut and marine bacterial communities}}.
\newblock {\emph{\JournalTitle{npj Biofilms and Microbiomes}}}
  \textbf{\bibinfo{volume}{6}}, \doiprefix\url{10.1038/s41522-020-00145-9}
  (\bibinfo{year}{2020}).

\bibitem{Shaw2019ModellingMR}
\bibinfo{author}{Shaw, L.~P.} \emph{et~al.}
\newblock \bibinfo{journal}{\bibinfo{title}{Modelling microbiome recovery after
  antibiotics using a stability landscape framework}}.
\newblock {\emph{\JournalTitle{The ISME Journal}}}
  \textbf{\bibinfo{volume}{13}}, \bibinfo{pages}{1845 -- 1856}
  (\bibinfo{year}{2019}).

\bibitem{Faust2015MetagenomicsMT}
\bibinfo{author}{Faust, K.}, \bibinfo{author}{Lahti, L.},
  \bibinfo{author}{Gonze, D.}, \bibinfo{author}{de~Vos, W.~M.} \&
  \bibinfo{author}{Raes, J.}
\newblock \bibinfo{journal}{\bibinfo{title}{Metagenomics meets time series
  analysis: unraveling microbial community dynamics.}}
\newblock {\emph{\JournalTitle{Current opinion in microbiology}}}
  \textbf{\bibinfo{volume}{25}}, \bibinfo{pages}{56--66}
  (\bibinfo{year}{2015}).

\bibitem{AttentionWhatYouNeed}
\bibinfo{author}{Vaswani, A.} \emph{et~al.}
\newblock \bibinfo{title}{Attention is all you need}.
\newblock In \emph{\bibinfo{booktitle}{Proceedings of the 31st International
  Conference on Neural Information Processing Systems}},
  \bibinfo{pages}{6000–6010}, \doiprefix\url{10.5555/3295222.3295349}
  (\bibinfo{year}{2017}).

\bibitem{graphkke}
\bibinfo{author}{Melnyk, K.}, \bibinfo{author}{Klus, S.},
  \bibinfo{author}{Montavon, G.} \& \bibinfo{author}{Conrad, T.~O.}
\newblock \bibinfo{journal}{\bibinfo{title}{Graphkke: graph kernel koopman
  embedding for human microbiome analysis.}}
\newblock {\emph{\JournalTitle{Applied Network Science}}}
  \textbf{\bibinfo{volume}{5}}, \doiprefix\url{10.1007/s41109-020-00339-2}
  (\bibinfo{year}{2020}).

\bibitem{CholeraInfOriginal}
\bibinfo{author}{Hsiao, A.} \emph{et~al.}
\newblock \bibinfo{journal}{\bibinfo{title}{Members of the human gut microbiota
  involved in recovery from vibrio cholerae infection}}.
\newblock {\emph{\JournalTitle{Nature}}} \textbf{\bibinfo{volume}{515}},
  \bibinfo{pages}{423--426}, \doiprefix\url{10.1038/nature13738}
  (\bibinfo{year}{2014}).

\bibitem{Kazemi2020RepresentationLF}
\bibinfo{author}{Kazemi, S.~M.} \emph{et~al.}
\newblock \bibinfo{journal}{\bibinfo{title}{Representation learning for dynamic
  graphs: A survey}}.
\newblock {\emph{\JournalTitle{J. Mach. Learn. Res.}}}
  \textbf{\bibinfo{volume}{21}}, \bibinfo{pages}{70:1--70:73}
  (\bibinfo{year}{2020}).

\bibitem{Barros2021ASO}
\bibinfo{author}{Barros, C. D.~T.}, \bibinfo{author}{Mendonça, M. R.~F.},
  \bibinfo{author}{Vieira, A.~B.} \& \bibinfo{author}{Ziviani, A.}
\newblock \bibinfo{journal}{\bibinfo{title}{A survey on embedding dynamic
  graphs}}.
\newblock {\emph{\JournalTitle{ArXiv}}}
  \textbf{\bibinfo{volume}{abs/2101.01229}} (\bibinfo{year}{2021}).

\bibitem{Cui2019ASO}
\bibinfo{author}{Cui, P.}, \bibinfo{author}{Wang, X.}, \bibinfo{author}{Pei,
  J.} \& \bibinfo{author}{Zhu, W.}
\newblock \bibinfo{journal}{\bibinfo{title}{A survey on network embedding}}.
\newblock {\emph{\JournalTitle{IEEE Transactions on Knowledge and Data
  Engineering}}} \textbf{\bibinfo{volume}{31}}, \bibinfo{pages}{833--852}
  (\bibinfo{year}{2019}).

\bibitem{Zhang2020NetworkRL}
\bibinfo{author}{Zhang, D.}, \bibinfo{author}{Yin, J.}, \bibinfo{author}{Zhu,
  X.} \& \bibinfo{author}{Zhang, C.}
\newblock \bibinfo{journal}{\bibinfo{title}{Network representation learning: A
  survey}}.
\newblock {\emph{\JournalTitle{IEEE Transactions on Big Data}}}
  \textbf{\bibinfo{volume}{6}}, \bibinfo{pages}{3--28} (\bibinfo{year}{2020}).

\bibitem{GraphRepSurvey2022}
\bibinfo{author}{Khoshraftar, S.} \& \bibinfo{author}{An, A.}
\newblock \bibinfo{title}{A survey on graph representation learning methods},
  \doiprefix\url{10.48550/ARXIV.2204.01855} (\bibinfo{year}{2022}).

\bibitem{node2vec}
\bibinfo{author}{Grover, A.} \& \bibinfo{author}{Leskovec, J.}
\newblock \bibinfo{title}{Node2vec: Scalable feature learning for networks}.
\newblock In \emph{\bibinfo{booktitle}{Proceedings of the 22nd ACM SIGKDD
  International Conference on Knowledge Discovery and Data Mining}},
  \bibinfo{pages}{855--864}, \doiprefix\url{10.1145/2939672.2939754}
  (\bibinfo{year}{2016}).

\bibitem{DeepWalk}
\bibinfo{author}{Perozzi, B.}, \bibinfo{author}{Al-Rfou, R.} \&
  \bibinfo{author}{Skiena, S.}
\newblock \bibinfo{title}{Deepwalk: Online learning of social representations}.
\newblock In \emph{\bibinfo{booktitle}{KDD '14: Proceedings of the 20th ACM
  SIGKDD international conference on Knowledge discovery and data mining}},
  \bibinfo{pages}{701--710}, \doiprefix\url{10.1145/2623330.2623732}
  (\bibinfo{year}{2014}).

\bibitem{graph2vec}
\bibinfo{author}{Narayanan, A.} \emph{et~al.}
\newblock \bibinfo{journal}{\bibinfo{title}{{graph2vec: Learning Distributed
  Representations of Graphs}}}.
\newblock {\emph{\JournalTitle{ArXiv}}} \doiprefix\url{10.1145/1235}
  (\bibinfo{year}{2017}).

\bibitem{Goyal2018DynGEMDE}
\bibinfo{author}{Goyal, P.}, \bibinfo{author}{Kamra, N.}, \bibinfo{author}{He,
  X.} \& \bibinfo{author}{Liu, Y.}
\newblock \bibinfo{journal}{\bibinfo{title}{Dyngem: Deep embedding method for
  dynamic graphs}}.
\newblock {\emph{\JournalTitle{ArXiv}}}
  \textbf{\bibinfo{volume}{abs/1805.11273}} (\bibinfo{year}{2018}).

\bibitem{dyngraph2vec}
\bibinfo{author}{Goyal, P.}, \bibinfo{author}{Rokka~Chhetri, S.} \&
  \bibinfo{author}{Canedo, A.}
\newblock \bibinfo{journal}{\bibinfo{title}{dyngraph2vec: Capturing network
  dynamics using dynamic graph representation learning}}.
\newblock {\emph{\JournalTitle{Knowl. Based Syst.}}}
  \textbf{\bibinfo{volume}{187}}, \doiprefix\url{10.1016/j.knosys.2019.06.024}
  (\bibinfo{year}{2020}).

\bibitem{EvolveGCN}
\bibinfo{author}{Pareja, A.} \emph{et~al.}
\newblock \bibinfo{title}{Evolvegcn: Evolving graph convolutional networks for
  dynamic graphs}, \doiprefix\url{10.48550/ARXIV.1902.10191}
  (\bibinfo{year}{2019}).

\bibitem{DySat}
\bibinfo{author}{Sankar, A.}, \bibinfo{author}{Wu, Y.}, \bibinfo{author}{Gou,
  L.}, \bibinfo{author}{Zhang, W.} \& \bibinfo{author}{Yang, H.}
\newblock \bibinfo{title}{Dysat: Deep neural representation learning on dynamic
  graphs via self-attention networks}.
\newblock In \emph{\bibinfo{booktitle}{Proceedings of the 13th International
  Conference on Web Search and Data Mining}}, \bibinfo{pages}{519--527}
  (\bibinfo{year}{2020}).

\bibitem{WL}
\bibinfo{author}{Shervashidze, N.}, \bibinfo{author}{Schweitzer, P.},
  \bibinfo{author}{van Leeuwen, E.~J.}, \bibinfo{author}{Mehlhorn, K.} \&
  \bibinfo{author}{Borgwardt, K.~M.}
\newblock \bibinfo{journal}{\bibinfo{title}{{Weisfeiler--{L}ehman Graph
  Kernels}}}.
\newblock {\emph{\JournalTitle{Journal of Machine Learning Research}}}
  \textbf{\bibinfo{volume}{12}}, \bibinfo{pages}{2539--2561}
  (\bibinfo{year}{2011}).

\bibitem{doc2vec}
\bibinfo{author}{Le, Q.} \& \bibinfo{author}{Mikolov, T.}
\newblock \bibinfo{title}{Distributed representations of sentences and
  documents}.
\newblock In \emph{\bibinfo{booktitle}{Proceedings of the 31st International
  Conference on International Conference on Machine Learning}},
  vol.~\bibinfo{volume}{32}, \doiprefix\url{10.5555/3044805.3045025}
  (\bibinfo{year}{2014}).

\bibitem{TransformerIB}
\bibinfo{author}{Chefer, H.}, \bibinfo{author}{Gur, S.} \&
  \bibinfo{author}{Wolf, L.}
\newblock \bibinfo{journal}{\bibinfo{title}{Transformer interpretability beyond
  attention visualization}}.
\newblock {\emph{\JournalTitle{2021 IEEE/CVF Conference on Computer Vision and
  Pattern Recognition (CVPR)}}} \bibinfo{pages}{782--791}
  (\bibinfo{year}{2021}).

\bibitem{LRP}
\bibinfo{author}{Bach, S.} \emph{et~al.}
\newblock \bibinfo{journal}{\bibinfo{title}{On pixel-wise explanations for
  non-linear classifier decisions by layer-wise relevance propagation}}.
\newblock {\emph{\JournalTitle{PLoS ONE}}} \textbf{\bibinfo{volume}{10(7)}}
  (\bibinfo{year}{2015}).

\bibitem{Langdon2016TheEO}
\bibinfo{author}{Langdon, A.~E.}, \bibinfo{author}{Crook, N.} \&
  \bibinfo{author}{Dantas, G.}
\newblock \bibinfo{journal}{\bibinfo{title}{The effects of antibiotics on the
  microbiome throughout development and alternative approaches for therapeutic
  modulation}}.
\newblock {\emph{\JournalTitle{Genome Medicine}}} \textbf{\bibinfo{volume}{8}}
  (\bibinfo{year}{2016}).

\bibitem{ViT2020}
\bibinfo{author}{Dosovitskiy, A.} \emph{et~al.}
\newblock \bibinfo{title}{An image is worth 16x16 words: Transformers for image
  recognition at scale}.
\newblock In \emph{\bibinfo{booktitle}{ICLR}} (\bibinfo{year}{2021}).

\bibitem{GraphAttentionNetwork}
\bibinfo{author}{Veli{\v{c}}kovi{\'{c}}, P.} \emph{et~al.}
\newblock \bibinfo{journal}{\bibinfo{title}{{Graph Attention Networks}}}.
\newblock {\emph{\JournalTitle{International Conference on Learning
  Representations}}}  (\bibinfo{year}{2018}).

\bibitem{GeneralizationOfTransformer}
\bibinfo{author}{Dwivedi, V.~P.} \& \bibinfo{author}{Bresson, X.}
\newblock \bibinfo{journal}{\bibinfo{title}{A generalization of transformer
  networks to graphs}}.
\newblock {\emph{\JournalTitle{AAAI Workshop on Deep Learning on Graphs:
  Methods and Applications}}}  (\bibinfo{year}{2021}).

\bibitem{SimpleContrastiveLearning}
\bibinfo{author}{Chen, T.}, \bibinfo{author}{Kornblith, S.},
  \bibinfo{author}{Norouzi, M.} \& \bibinfo{author}{Hinton, G.}
\newblock \bibinfo{title}{A simple framework for contrastive learning of visual
  representations}.
\newblock In \emph{\bibinfo{booktitle}{Proceedings of the 37th International
  Conference on Machine Learning}}, vol. \bibinfo{volume}{119},
  \bibinfo{pages}{1597--1607} (\bibinfo{year}{2020}).

\bibitem{gehring17a}
\bibinfo{author}{Gehring, J.}, \bibinfo{author}{Auli, M.},
  \bibinfo{author}{Grangier, D.}, \bibinfo{author}{Yarats, D.} \&
  \bibinfo{author}{Dauphin, Y.~N.}
\newblock \bibinfo{title}{Convolutional sequence to sequence learning}.
\newblock In \bibinfo{editor}{Precup, D.} \& \bibinfo{editor}{Teh, Y.~W.}
  (eds.) \emph{\bibinfo{booktitle}{Proceedings of the 34th International
  Conference on Machine Learning}}, vol.~\bibinfo{volume}{70} of
  \emph{\bibinfo{series}{Proceedings of Machine Learning Research}},
  \bibinfo{pages}{1243--1252} (\bibinfo{publisher}{PMLR},
  \bibinfo{year}{2017}).

\end{thebibliography}



\section*{Acknowledgements}
This work was supported by the Forschungscampus MODAL (project grant 3FO18501) and funded by Berlin Institute for the Foundations of Learning and Data (BIFOLD, ref. 01IS18025A and ref. 01IS18037I) and by the Deutsche Forschungsgemeinschaft (DFG, German Research Foundation) under Germany Excellence Strategy – The Berlin Mathematics Research Center MATH+ (EXC-2046/1, project ID: 390685689).

\section*{Author contributions statement}
TC designed and supervised the research; KW designed the architecture of the model; KM performed the research, wrote the code and the manuscript; TC and KW revised and corrected the manuscript content. All authors read and approved the final manuscript.

\section*{Additional information}
\textbf{Accession codes} Code can be found here: \url{https://github.com/k-melnyk/deep-metastability}; 
\textbf{Competing interests} The authors declare no competing interests.

\end{document}